\documentclass[twocolumn,pre]{revtex4}

\usepackage{url}

\usepackage{amsmath}
\usepackage{graphicx}
\usepackage{subfigure}

\renewcommand \thesubsection{\arabic{section}.\arabic{subsection}}

\newcommand{\simlt}
 {\ifmmode       { \raisebox{-.4em}{$<$}\atop\sim}
    \else        {$\raisebox{-.4em}{$<$}\atop\sim$}
 \fi}

\begin{document}

\title{Satisfiability, sequence niches, and molecular codes in 
cellular signaling}
\author{Christopher R. Myers}
\affiliation{Computational Biology Service Unit,
Center for Advanced Computing and 
Life Sciences Core Laboratories Center,
Cornell University, Ithaca, NY, USA}

\begin{abstract}

Biological information processing as implemented by regulatory and
signaling networks in living cells requires sufficient specificity of
molecular interaction to distinguish signals from one another, but
much of regulation and signaling involves somewhat fuzzy and
promiscuous recognition of molecular sequences and structures, which
can leave systems vulnerable to crosstalk.  This paper examines a
simple computational model of protein-protein interactions which
reveals both a sharp onset of crosstalk and a fragmentation of the
neutral network of viable solutions as more proteins compete for
regions of sequence space, revealing intrinsic limits to reliable
signaling in the face of promiscuity.  These results suggest
connections to both phase transitions in constraint satisfaction
problems and coding theory bounds on the size of communication codes.

\end{abstract}

\maketitle

\section{Introduction}
\label{Sec-Introduction}

The functioning of complex biomolecular pathways hinges on conveying
molecular signals reliably in the stochastic and evolving milieu of
living cells.  These signals are mediated by molecular interactions
that distinguish physiological binding partners from myriad other
cellular constituents: this ability to distinguish functional signals
from the molecular noise is ultimately the source of information
processing in cellular networks.  But molecular recognition is subtle:
many of the molecular interactions involved in cellular regulatory and
signaling pathways do not involve highly specific ``lock and key''
binding, but instead are characterized by more fuzzy and promiscuous
recognition of families of sequences and configurations
\cite{Ptashne2002,Mayer2001, Castagnoli2004}. Furthermore, there are
often multiple types of molecules within a cell that can bind to the
same target, such as different proteins containing homologous copies
of a modular interaction domain.  We therefore ask a basic theoretical
question concerning cellular signaling in crowded sequence spaces,
where multiple proteins bind to similar families of molecular
sequences and structures: under what circumstances can crosstalk be
avoided in such a system?  This paper investigates a simple null
model, associated with random molecular sequences, that is amenable to
analysis and suggests connections to recent work on phase transitions
in combinatorial NP-complete problems.  This random model is not
directly applicable to the evolved molecular sequences found in
nature, but serves as a useful first step in defining the landscape of
constraint satisfaction in cellular signaling.

The theory of communication in noisy channels, dating back to the
seminal work of Shannon \cite{Shannon1948, Shannon1949}, also provides
a useful framework in which to interpret cellular signals.  Engineered
error-correcting codes embed messages in higher-dimensional spaces
(e.g., via encoded checks on the message integrity), to insulate each
possible codeword within a sphere in the embedding space.  By packing
such spheres so that they are disjoint, any corrupted message word can
(up to some defined number of errors) be uniquely associated with an
original code word.  In molecular signaling, \emph{sequence
recognition volumes} play a similar role: these volumes describe the
sets of sequences recognized (i.e., bound with significant
probability) by various target molecules.  In molecular signaling,
overlapping recognition of sequences precludes the sort of disjoint
geometries found in engineered codes.  Instead of asking, therefore,
whether \emph{all} messages can be communicated through a protein
signaling channel, we will focus here instead on whether \emph{any}
message can be so conveyed (under the assumption that evolutionary
selection might find such a solution if it does in principle exist).
Addressing the discrimination of potentially ambiguous signals, this
work is related to issues arising in error-correcting codes, but
geometrically it is in some ways more similar to problems involving
covering codes \cite{Cohen1997} and identifying codes
\cite{Karpovsky1998}.  A central result presented here, which
establishes limits on the number of proteins that can compete for
regions in sequence space before crosstalk becomes likely, is akin to
a bound on the size of a code in a communication system.

This work was motivated in part by experiments on SH3-mediated
signaling in yeast (\emph{Saccharomyces cerevisiae}), by Zarrinpar,
Park and Lim \cite{Zarrinpar2003}.  SH3 domains constitute a family of
conserved modular protein domains, known to bind to a set of
proline-rich peptide sequences (the so-called ``PXXP'' motif, which
actually consists of a larger peptide of approximately 8-10
residues)\cite{Mayer2001, Cesareni2002}. Because of this interaction
promiscuity, and because several proteins in yeast contain SH3
domains, it was not obvious whether there would be crosstalk among
pathways involving different SH3-containing proteins.  Zarrinpar
\emph{et al}. probed the yeast high-osmolarity signaling pathway,
which involves the interaction of Sho1 (a protein with an SH3 domain)
and Pbs2 (containing a PXXP motif).  By making chimeric versions of
Sho1 containing different SH3 domains, they demonstrated that none of
the other native yeast SH3 domains were capable of interacting with
Pbs2, but that half of the metazoan SH3 domains they tested were able
to do so.  They surmised that there has been an evolutionary selection
against crosstalk with that pathway in yeast, with protein sequences
having co-evolved such that the Pbs2 ligand lies in a niche in
sequence space where it is recognized by only the Sho1 SH3 domain.
Since there has been no such selection pressure to avoid crosstalk in
other organisms, the Pbs2 motif bound to non-native SH3 domains with
greater probability.  (See supplementary text and Figure S.1 for
further discussion.)  It is the structure of these sorts of
\emph{sequence niches} that form the core of this paper.  In related
work, Sear has computed the capability of a set of competitive
protein-protein interactions, and examined crosstalk avoidance in a
model motivated by the same set of yeast signaling experiments
\cite{Sear2004a, Sear2004b}.

The fundamental questions posed by the experiments on SH3 signaling in
yeast extend beyond that particular system.  A classic problem in
immunology is the ability of antibodies to discriminate between
``self'' and ``nonself'' antigens, and early work addressed how large
a recognition region needs to be in order to reliably perform this
discrimination \cite{Percus1993}.  In gene regulation, transcription
factors (TFs) that regulate gene expression by binding to DNA are
organized in families that often recognize similar sorts of sequences.
Recent work in that area has explored tradeoffs between binding
specificity and system robustness \cite{Sengupta2002}, balances
between selection and mutation \cite{Gerland2002a}, evolutionary
divergence of competing TF-binding sequence pairs to avoid crosstalk
\cite{Poelwijk2006}, and the application of ideas from coding theory
to understand limits on the size of TF families \cite{Itzkovitz2006}.
In bacterial signaling, the possibility of crosstalk among
two-component regulatory systems, whereby multiple response regulators
are activated by a single sensor kinase, has also been explored to
gain insight into how environmental signals are combined
\cite{Bijlsma2003,Hellingwerf2005,Laub2007}.

\section{Results}
\label{Sec-Results}

\subsection{The Sequence Niche Question}
\label{Subsec-SequenceNiche}

\begin{figure}[t!]
\centering
\includegraphics[width=65mm]{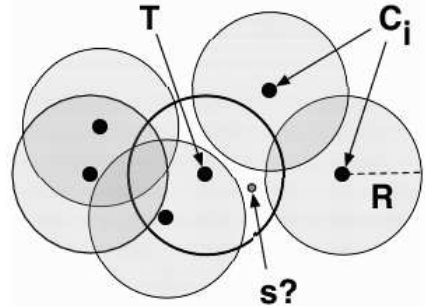}
\caption{\label{Fig_seqniche}
The Sequence Niche Question: 
given a target protein sequence 
$T$ and a set of $N$ crosstalking protein sequences $\{C\}$, 
is there a sequence $s$ that is bound by $T$ but not by any 
of the proteins $C_i$?  In this model, 
sequences are binary strings of length $L$,
and two sequences bind if the Hamming distance between them 
is less than or equal to $R$.
}
\end{figure}

We begin by distilling the central question to be considered here:
under what conditions does a unique sequence
niche exist so that signaling without crosstalk might be possible?  To
address this question, we adopt a highly abstracted model of
protein-protein interaction, in which protein sequences are
represented by binary strings of length $L$ (consisting of $0$'s and $1$'s)
rather than as peptide strings in the 20-letter amino acid alphabet.
(Binary sequence models, such as the HP model,
has been used in the study of protein folding \cite{Lau1989}, 
although it remains an open question as to 
whether there is an appropriate coarse-grained alphabet capable of 
capturing the essential biochemistry of protein-protein interactions 
involved in signaling.)
In this model, binding of a peptide sequence to a protein is achieved if
the sequence is sufficiently close to the consensus sequence
recognized by the protein, with Hamming distance
used as a measure of closeness: two sequences bind if they differ in
at most $R$ positions, given some promiscuity radius $R$.  
Given this representation, we can pose the Sequence Niche Question,
phrased and typeset in the canonical style of Garey and
Johnson \cite{Garey1979} and illustrated schematically in
Fig.~\ref{Fig_seqniche}:\\[1pt]

\noindent{\bf SEQUENCE NICHE}\\[0pt]
INSTANCE: Binary sequence $T$ of length $L$, 
a set of binary crosstalk sequences $C_i$, for $i=1,...,N$, 
each of length $L$, 
and an integer $R$, $0 \le R \le L$.\\[0pt]
QUESTION: Is there a binary sequence $s$ of length $L$
such that $H(T,s) \le R$ 
and $H(C_i,s) > R$ for $i=1,..,N$, where
$H(x,y)$ is the Hamming distance between sequences $x$ and $y$?\\[1pt]

\begin{figure*}[t!]
\centering
\includegraphics[width=180mm]{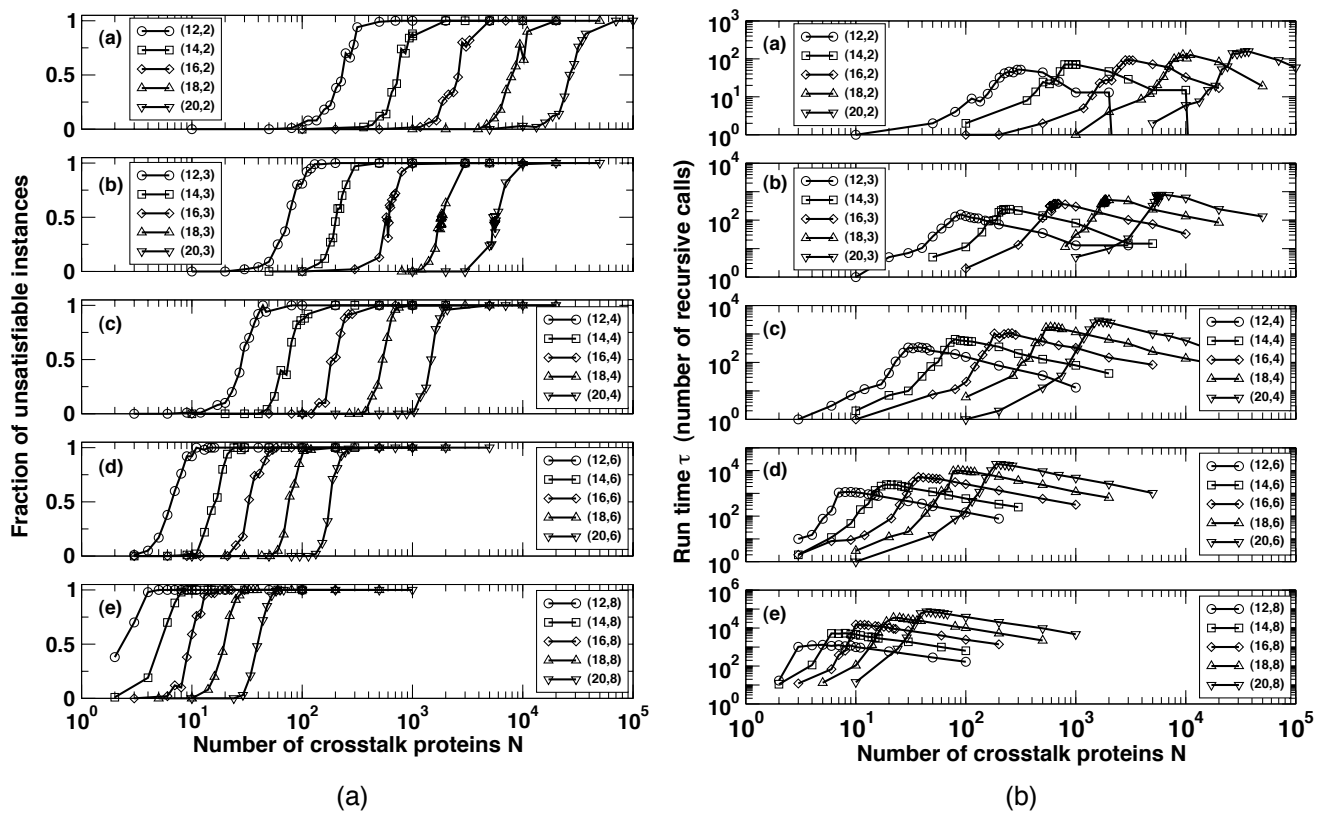}
\caption{\label{Fig_SATFrac}
(a) Average fraction of unsatisfiable instances of the 
random SNQ as a function of $L$,$R$, and $N$ ($(L,R)$ specified in 
figure legend, $N$ varying along x-axis).
(b) Average run time $\tau$ of the SNQ decision 
(number of recursive calls in the solution algorithm) 
for the same instances depicted in (a).  
Averages are over 100 instances of the SNQ for each $(L,R,N)$ set.
}
\end{figure*}

The Sequence Niche Question (SNQ) is a rephrasing of the
Distinguishing String Selection Problem (DSSP), as defined by Lanctot
\emph{et al.} \cite{Lanctot2003}.  (The DSSP allows more for
multiple ``good'' strings $S_c$ to be matched within some Hamming
distance $k_c$ and multiple ``bad'' strings $S_f$ to be avoided
outside some Hamming distance $k_f$.)  The DSSP was proven to be
NP-complete \cite{Lanctot2003}; the SNQ is the DSSP with $S_c=1$ and
$k_c=k_f$, but the computational complexity of the DSSP does not
depend on the values of these parameters, so the SNQ is also
NP-complete.  The SNQ is similar in spirit to the well-known computer
science problem SAT (and its specialization $k-$SAT), in that these
problems ask whether there exists a solution that satisfies a set of
(potentially conflicting) constraints \cite{Garey1979}.  Borrowing
from the language of SAT, we say a particular instance of the SNQ is
``satisfiable'' when a solution $s$ exists, and ``unsatisfiable'' when
there is no such solution.  The SNQ asks whether discrimination of one
target protein from a background of crosstalking proteins is possible.
A symmetric generalization of this problem would ascertain whether
every protein in a collection is distinguishable, that is, whether
there is a separate sequence niche for each of the $N$ proteins.  This
generalized SNQ is essentially that considered by Sear
\cite{Sear2004b}, although he did so for a 4-letter protein alphabet
and a more realistic treatment of the binding kinetics than simple
Hamming distances, demonstrating that for at least some parameters,
such discrimination is possible.  The generalized SNQ is presumably in
the same complexity class as the single-target SNQ, since deciding it
simply involves deciding $N$ separate SNQs.

\subsection{Satisfiability of the random SNQ}

The NP-completeness of the SNQ is a statement about its worst-case
complexity, but there has been increasing interest in recent years in
quantifying the typical-case complexity of NP-hard problems.  A common
strategy is to examine ensembles of random instances of NP-hard
problems, investigating how solution complexity depends upon
parameters that characterize those random instances.  A similar
strategy is adopted here.

Multiple random instances of the SNQ were examined (with uniform equal 
probability of $0$'s and $1$'s in the sequence strings),
for various values of the problem parameters $L$, $R$, and $N$.
Figure~\ref{Fig_SATFrac}(a) shows the average unsatisfiable fraction
of random SNQ instances as a function of the number of crosstalking
proteins $N$, averaged over an ensemble of 100 random instances for each 
$N$. In addition, Figure~\ref{Fig_SATFrac}(b) shows the average run
time $\tau$ required for determining whether or not an instance is
satisfiable (where run time is measured in units of the number of
recursive calls to the solution algorithm of Gramm \emph{et al.}
\cite{Gramm2003}).  Fig.~\ref{Fig_SATFrac}(a) demonstrates a
transition from satisfiability (SAT) to unsatisfiability (UNSAT) as
the number of crosstalking proteins is increased.  Rather than a
gradual diminution in the capacity for reliable signaling, the SNQ
exhibits a relatively abrupt switch as $\log N$ increases.
Fig.~\ref{Fig_SATFrac}(b) reveals, for the same set of parameter
values, that the run time of the solution algorithm reaches a maximum
near the point of the SAT-UNSAT transition.  In other words, it becomes
significantly more difficult to decide if a given instance is
satisfiable or not when that instance lies near the transition.  The
characteristic scales of the random SNQ are seen to vary over orders
of magnitude.  For the solution run times, this is perhaps not surprising:
since the SNQ is NP-complete, we expect the worst-case run time of the
solution algorithm to be exponential in the size of the problem.

\subsection{Scaling of the SNQ transition: a satisfiability bound on 
the number of crosstalking proteins}

\begin{figure}[t!]
\centering
\includegraphics[width=85mm]{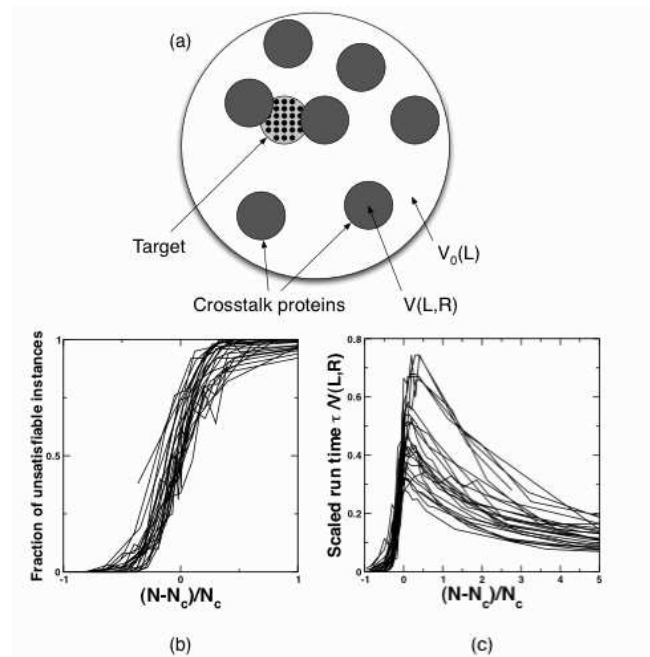}
\caption{\label{Fig_scaled_SATFrac}
Scaling description of the SAT-UNSAT transition in the SNQ.
(a) Schematic depiction of the covering of available sequences (black dots)
in the target volume as crosstalk proteins (gray circles) 
are laid down randomly.
(b, c) Scaling of the satisfiability and run time data 
in Fig.~\ref{Fig_SATFrac}
based on the scaling theory presented: (b) the number of crosstalk proteins $N$
are scaled by $N \to (N-N_c)/N_c$, and (c) in addition to scaling $N$,
the run times $\tau$ are
scaled by the number of sequences in the target 
volume $V(L,R)$ that must be considered.  
}
\end{figure}

Even though the characteristic scales of the SNQ vary by
orders of magnitude, there is a scaling structure evident in those data.
This structure is understood by considering the geometric and
probabilistic nature of the random SNQ.  A given 
instance is unsatisfiable if the target volume (i.e., the Hamming sphere 
of radius $R$ surrounding the target sequence $T$) is completely covered
by the union of the crosstalk volumes (centered about the crosstalk
sequences $\{C\}$), a process that is illustrated 
schematically in Fig.~\ref{Fig_scaled_SATFrac}(a).
We can estimate the critical number of crosstalk
proteins $N_c$ needed to cover the sequence volume of the 
target protein (see supplementary text for full derivation):
\begin{equation}
N_c =
\frac{{\log(1/V)}}{{\log(1-V/V_0)}} \label{eq:Nc}
\end{equation}
where $V_0(L) = 2^L$ is the total number of possible binary 
sequences of length $L$,
and $V(L,R) = \sum_{n=0}^R {L\choose{n}}$ is the 
number of binary sequences in a ball 
of Hamming radius $R$ about a given sequence.
We can interpret this as a \emph{random satisfiability bound} 
on the approximate number of randomly distributed proteins that 
can coexist without crosstalk.

With this critical protein number, we can rescale the
raw satisfiability and run time data of Fig.~\ref{Fig_SATFrac}.  These
rescaled data are shown in Fig.~\ref{Fig_scaled_SATFrac}; in (b) and
(c) the protein number (x-axis) is scaled as $N \to (N-N_c)/N_c$, and
in (c), the run time data (y-axis) are scaled by the exponentially
growing number of sequences in the search tree $V(L,R)$ that in
principle need to be considered.  
The collapse of each set of unscaled data onto a reasonably compact 
scaling form suggests this simple description is correct.  

\subsection{Fragmentation of the solution space}

Previously we considered whether 
there is \emph{any} solution to a given instance of the SNQ.  
Here we examine the structure of the space of \emph{all} satisfying 
solutions for an instance, as determined via exhaustive enumeration.

\begin{figure*}[t!]
\begin{center}
\subfigure[]{
\includegraphics[width=42mm]{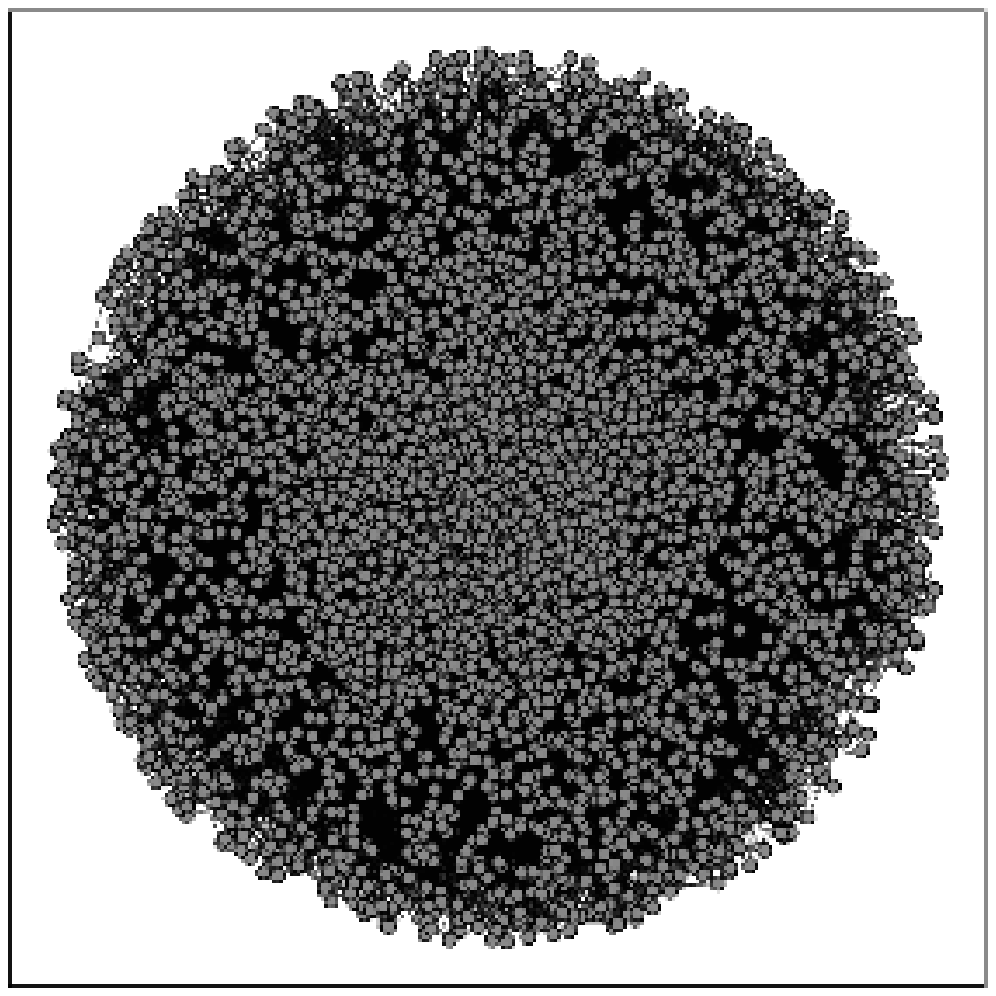}}
\subfigure[]{
\includegraphics[width=42mm]{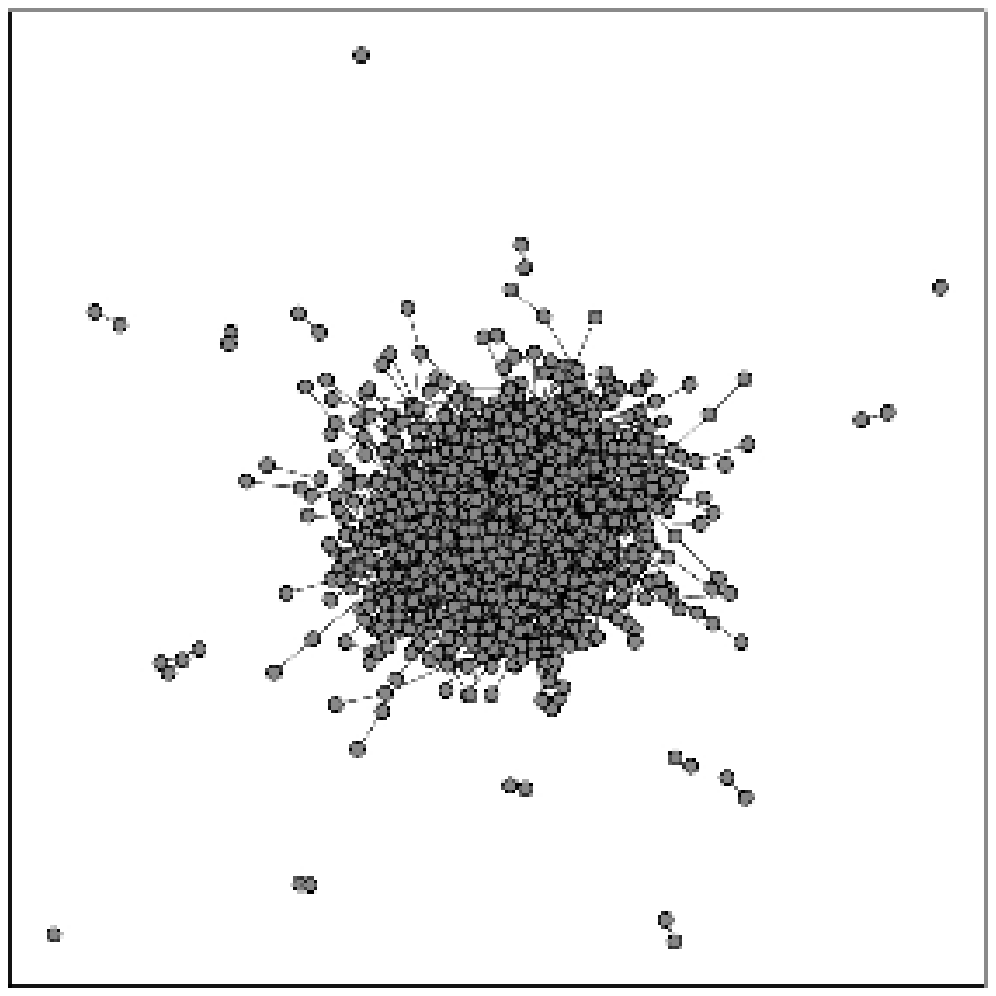}}
\subfigure[]{
\includegraphics[width=42mm]{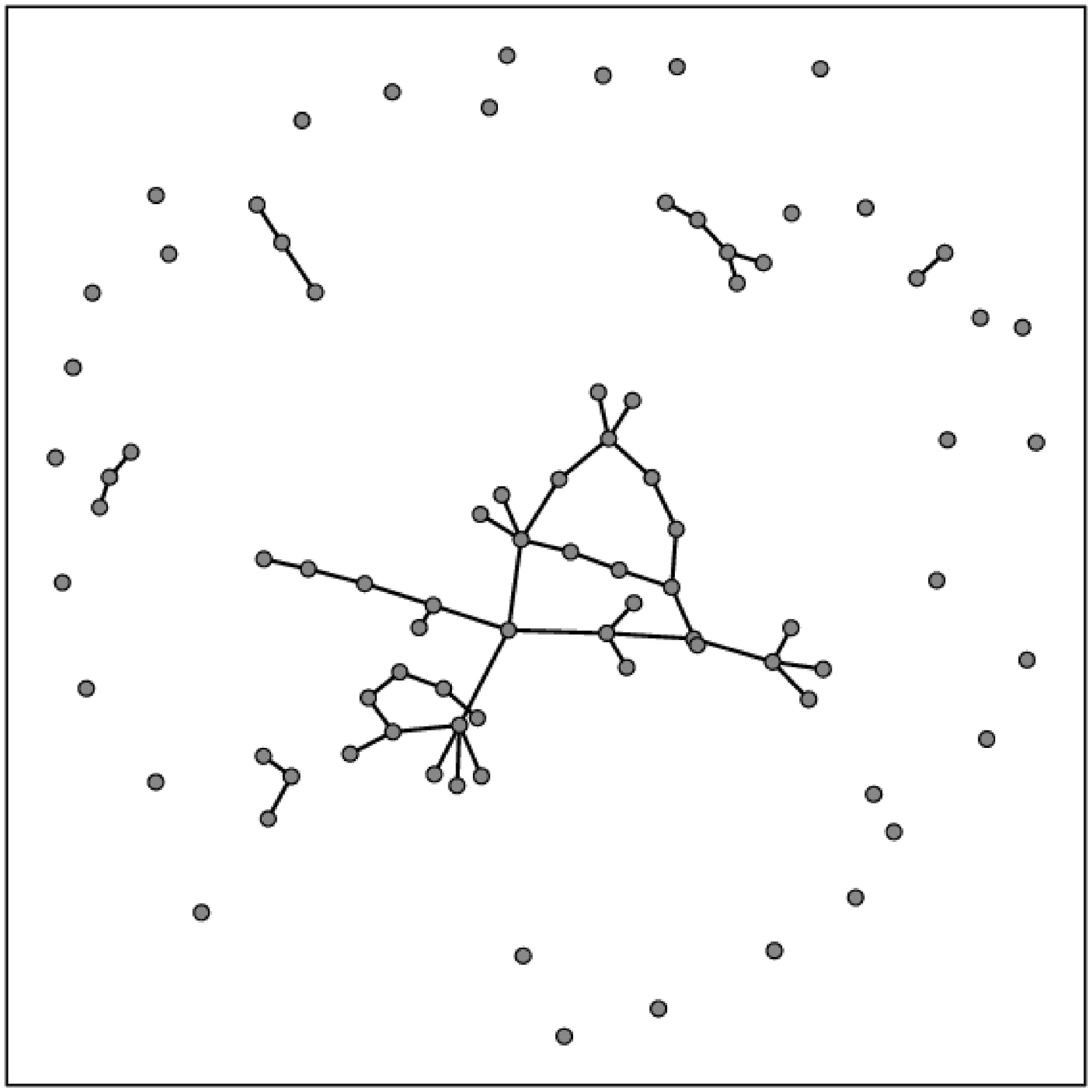}}
\subfigure[]{
\includegraphics[width=39mm]{cluster_averages_16_6_v2.eps}}
\caption{\label{Fig_fragmentation}
Fragmentation of the solution space as the 
SAT-UNSAT transition is approached.
(a, b, c) The neutral network 
of satisfying solutions 
$\{s_N\}$ for one particular problem instance ($L=16, R=6$), 
as a function of number of crosstalking proteins $N$.
Satisfying sequences
(nodes) are connected by edges (lines) in a network if they 
are separated by Hamming distance 1.  (The spatial layout of nodes has 
no meaning; all sequences are vertices on an $L-$dimensional hypercube.)
(a) $N=4$: there are 5786 satisfying solutions in one large 
connected component.  This cluster is broken up into multiple pieces 
as $N$ increases. (b) $N=12$: 1226 sequences are distributed among
18 connected components.  (c) $N=20$: only 85 sequences remain viable,
scattered across 38 disjoint components.
(d) For $L=16, R=6$, average values 
of the size of the largest connected 
sequence cluster (solid line) and the number of disjoint clusters (dashed 
line) as a function of $N$, averaged over 100 SNQ instances for each value 
of $N$.
}
\end{center}
\end{figure*}

Consider a fixed target sequence $T$ and a set of potential crosstalk
sequences $\{C\}$.  Imagine introducing crosstalk sequences one at a
time, and identifying the set of all sequences $\{s_N\}$ that satisfy
the SNQ for that instance with $N$ crosstalk sequences.  Of particular
interest here is the size and structure of the solution set $\{s_N\}$
as a function of the number of proteins $N$.  For each set, we
assemble a graph whose nodes are sequences $s$ that satisfy the SNQ
and whose edges connect satisfying sequences if they are neighbors on
the hypercube, i.e., if their Hamming distance from each other is 1.
This graph represents the \emph{neutral network} of all solutions to a
\emph{given} instance of the SNQ, along which single point mutations
to the solution string (bit flips) can be made without producing
crosstalk.  For various $N$, we can compute the set of connected
components of the resulting graph.  The change in the structure of the
neutral network of satisfying solutions is illustrated, for a given
problem instance with $L=16$ and $R=6$, in
Fig.~\ref{Fig_fragmentation}.  For small numbers of proteins
(Fig.~\ref{Fig_fragmentation}(a)), there are many possible solutions
to the SNQ, and those solutions all coalesce into one connected
cluster, such that any solution can be reached from any other via a
succession of single-bit flips to the solution string.  As $N$
increases (Fig.~\ref{Fig_fragmentation}(b)), the number of satisfying
solutions decreases, and the connected cluster of solutions is
fragmented into many disjoint sets (still dominated by a central
core).  This fragmentation and evaporation of the sequence clusters
continues for larger $N$ (Fig.~\ref{Fig_fragmentation}(c)), until
finally all solutions disappear, and unique signaling is no longer
possible.  While the neutral networks shown reveal the effects of
mutations in the solution string $s$, it should be noted that single
point mutations in the sequences representing the centers of the
proteins $T$ and $\{C\}$ can result in drastic changes in the neutral
network topology, e.g., by fragmenting a single large cluster into a
set of smaller ones.

A summary of these trends is shown in Fig.~\ref{Fig_fragmentation}(d),
by averaging over many SNQ instances (for $L=16$ and $R=6$).  This
reveals that the size (i.e., the number of nodes) of the largest
cluster (solid line) decreases roughly exponentially with crosstalk
number $N$.  We can understand this decrease in part by considering
the geometric argument summarized in Fig.~\ref{Fig_scaled_SATFrac}(a);
see the supplementary text for details.  Also shown in
Fig.~\ref{Fig_fragmentation}(d) is the number of disjoint clusters
(dashed line); this is seen to initially increase with $N$ -- as the
single satisfying solution cluster is fragmented -- and then decrease
-- as small sequence clusters evaporate in the presence of new
crosstalk proteins.  Fig.~\ref{Fig_fragmentation} reveals a number of
isolated clusters of size $1$, but these problem sizes are rather
small (given the computational burdens of exhaustive enumeration).  It
is an open question whether nontrivial cluster size distributions will
reveal themselves as larger problem sizes are considered.

\section{Discussion}
\label{Sec-Discussion}

The goal of this paper has been to examine the limitations of 
crosstalk-free signaling in a simple model of competitive
protein-protein interactions, as a first step toward developing a more
comprehensive and realistic theory.  
The numerical experiments presented were motivated by
phase transitions observed in the random $k-$SAT problem
\cite{Mitchell1992, Kirkpatrick1994, Monasson1999, Friedgut1999}, 
where there is a SAT-UNSAT transition as the ratio of constraints to
variables is increased.  The numerical results presented for the SNQ
demonstrate something similar: a relatively sharp transition from
satisfiability to unsatisfiability with increasing competition for
sequence space, along with an increase in computational complexity
near the transition.  Phase transitions have been studied in a number
of NP-hard problems, although applications to biological problems have
been scant and generally at coarser levels of biological description
\cite{Correale2006,Gravner2007}, despite significant interest in the
computational complexity of problems involving sequence matching and
discrimination \cite{Lanctot2003, Gramm2003}.  A second phase
transition has more recently been identified in $k-$SAT, lurking near
the SAT-UNSAT phase boundary, involving the fragmentation of the set
of satisfying solutions \cite{Mezard2002, Mezard2003,Mezard2005}.  We
find evidence for such a fragmentation transition in small instances
of the SNQ, although further theoretical and computational work is
needed to fully characterize these transitions, which are only
strictly defined in the limit of infinite system size.

Of particular interest are the possible biological implications of
these results.  Where, for example, are signaling
systems in nature situated with respect to these types of phase
boundaries, and what sorts of codes has evolution uncovered 
in such systems?  Have evolutionary innovations -- such as novel interaction
domains \cite{Itzkovitz2006}, or scaffolds that localize 
signaling proteins \cite{Pawson1997}
-- arisen to rescue cellular networks from the precipice of crosstalk?  
Signaling interactions do not occur in isolation, but often involve
compartmentalization or localization (e.g., on scaffolds) that confer
context-dependent specificity in addition to the intrinsic sequence
specificity addressed here
\cite{Burack2000,Morrison2003,McClean2007}.  
In addition, fragmentation of the network of satisfying solutions of
the sort demonstrated here leads to complex neutral network
topologies.  The extent to which neutral network topology influences
evolution remains an open question \cite{Nimwegen1999,
Ciliberti2007a}.  Neutral network fragmentation could lead to
biological systems becoming frozen in local regions of sequence space,
unable to mutate to other satisfactory configurations far away.  This
could produce a sort of speciation at the molecular scale, perhaps
shedding light on phylogenetic relationships among related protein
interaction domains.  Larger-scale genomic rearrangements, such as
homologous recombination and horizontal transfer, may play a role in
helping biological communication systems become unstuck from a glassy,
fragmented phase where single point mutations are unable to do so.

Examination of the SAT-UNSAT transition in random instances of the SNQ
led to the derivation of a random satisfiability bound
(eq. (\ref{eq:Nc})).  This represents an upper limit to the number of
randomly distributed sequences that can coexist without crosstalk
becoming likely.  While the bound was motivated by the SAT-UNSAT
transition, it is also usefully interpreted within the context of
coding theory bounds on the size of codes.  Whereas a sphere-packing
bound \cite{Shannon1949} describes the number of Hamming spheres of
radius $R$ that can be packed in $L$ dimensions with no overlap, and a
smooth coding bound \cite{Itzkovitz2006} allows for some overlap of
sequence recognition spheres, our satisfiability bound is applicable
to a dense, overpacked limit when all capacity for uniquely
distinguishing signals disappears.  The bound presented in eq.\
(\ref{eq:Nc}) is explicitly applicable to binary sequences without
reverse-complement symmetry.  It is straightforwardly generalizable
(see supplementary text), within the assumption that binding is
entirely dictated by the Hamming distance between two sequences, to
sequences with larger alphabets (e.g., 20 amino acids) or to sequences
with reverse-complement symmetry (e.g., as has been done for other
code bounds treating DNA sequences \cite{Marathe2001, Itzkovitz2006}).

Protein sequences and sequence niches involved in cellular signaling
have, of course, been sculpted by evolution.  We might expect evolution
to be able to find better encoding schemes than the random placement considered
here, by arranging sequence recognition volumes to maximize fitness.
Addressing this question, however, requires consideration of several
factors.  First, it is not obvious what fitness measure is optimized
by natural selection.  If discrimination among different sequences
were the only determinant of fitness, we might expect encodings to
more closely resemble sphere packings, with recognition volumes
maximally distinct from one another.
Other determinants could alter such packings, however; a
fitness advantage from some weak crosstalk, perhaps as a form of
degeneracy or functional redundancy \cite{Edelman2001}, 
might keep recognition volumes from diverging
too far from one another.  And it must be remembered that evolutionary
mutation plays a central role in posing these constraint satisfaction
problems in the first place, in that gene duplication leads to the
creation of homologous proteins that recognize similar substrates.
The random limit considered here, while useful for analysis, is not
directly relevant to the biology of duplicated proteins that may diverge
from one another just far enough to be distinguishable \cite{Poelwijk2006}.

An examination of experimental and genomic data for model systems is
an obvious next step, both to probe the structure and evolution of
sequence niches in nature, and to develop more realistic and
predictive models of protein-protein interaction.  The experimental
work reported in ref. \cite{Zarrinpar2003} included a series of
single-base-pair missense mutations to the native yeast Pbs2 motif, to
probe the sequence space around that motif.  All such mutations
resulted in an increased cross-reactivity with other yeast SH3
domains, suggesting that the Pbs2 ligand lies near the periphery of a
possible sparse and tenuous sequence niche, but further examination of
yeast SH3 interaction data is needed to better characterize that.
Fortunately, there has been considerable experimental work in
screening synthetic peptide ligands to map out the sequence
recognition volumes of SH3 domains in several proteins
\cite{Landgraf2004}, and similar sorts of data are becoming available
for systems such as two-component regulators \cite{Laub2007}.
Computational classifiers (e.g., weight matrices and neural networks)
trained on protein-protein interaction data have been used to make
predictions about binding affinities of particular proteins to
arbitrary peptides
\cite{Brannetti2000,Ferraro2006}.  With a combination of experimental 
data, computational models, and a theoretical understanding of
the complexities of 
constraint satisfaction problems, we can aim to map out the structure
of high-dimensional sequence niches underlying cellular decision
making in biological systems.

\section{Acknowledgments}

This work was supported by USDA-ARS project 1907-21000-027-03.  I
would like to thank Jim Sethna, Bart Selman, Carla Gomes, Walter
Fontana, Marc M\'ezard, Sue Coppersmith, Chris Henley, Bistra Dilkina,
David Schneider, David Krakauer, and Bill Bialek for useful input.

\appendix

\numberwithin{figure}{section}
\setcounter{section}{19}     % S the 19th letter in the alphabet
\setcounter{equation}{0}
\setcounter{figure}{0}

\renewcommand \thesubsection{S.\arabic{subsection}}

\section*{Supplementary material}

\subsection{Derivation of critical number of crosstalking proteins (random 
satisfiability bound)}

Here we derive the result stated in eq. (1) of the main text, 
the critical number of 
crosstalking proteins $N_c$ for a given sequence length $L$ and 
promiscuity radius $R$, which we can interpret as a random satisfiability 
bound for the size of the protein-protein interaction code.
A given instance of the SNQ
is unsatisfiable if the target volume (i.e., the Hamming sphere 
of radius $R$ surrounding the target sequence $T$) is completely covered
by the union of the crosstalk volumes (centered about the crosstalk
sequences $\{C\}$), a process that is illustrated 
schematically in the main text in Fig. 3(a).
We can estimate the critical number of crosstalk
proteins $N_c$ needed to cover the sequence volume of the 
target protein.
For a given binary string of length $L$, the number of 
sequences $V(L,R)$ in a ball of Hamming radius $R$ is
\begin{equation}
V(L,R) = \sum_{n=0}^R {L\choose{n}}  \label{eq:VLR}
\end{equation}
and the total possible number of sequences $V_0(L)$ is 
\begin{equation}
V_0(L) = 2^L   \label{eq:V0L}
\end{equation}
Let $q$ be the ratio of these sequence volumes: 
\begin{equation}
q \equiv V/V_0 \label{eq:def_q}
\end{equation}
We consider depositing at random sequence volumes of size $V(L,R)$ in a
space of volume $V_0(L)$.  From the binomial distribution, the probability
that a given point in sequence space is covered $n$ times after $N$ 
proteins have been deposited is 
\begin{equation}
P_q(n|N) = {N\choose{n}} q^n (1-q)^{N-n} \label{eq:P_qnN}
\end{equation}
Therefore the probability $U_q(N)$ that a given point in
sequence space is left \emph{uncovered} by $N$ proteins 
is 
\begin{equation}
U_q(N) = P_q(0|N) = (1-q)^N \label{eq:U_qN}
\end{equation}
We can thus estimate the average number of 
sequences $S_u(V,q,N)$ in the target volume $V$ left uncovered 
by $N$ proteins to be
\begin{equation}
S_u(V,q,N) = V (1-q)^N  \label{eq:S_uVqN}
\end{equation}
We wish to 
estimate the critical number of proteins $N_c$ required to cover the 
target volume; since the sequence space is discrete, 
we estimate $N_c$ as the number
of proteins for which there is $O(1)$ remaining uncovered sequence
in the target volume. 
This yields 
\begin{equation}
V (1-q)^{N_c} = 1
\end{equation}
which implies
\begin{equation}
N_c =
\frac{{\log(1/V)}}{{\log(1-V/V_0)}} \label{eq:Nc}
\end{equation}

The estimate (\ref{eq:Nc}) 
appears to adequately describe the SNQ simulation data 
presented in the main text, as indicated by the scaling collapses 
shown in Fig. 3 of the main text.  
We expect the quality of the estimate to degrade, 
however, as the discrete nature of the sequence space becomes 
more important, i.e., 
as the number of sequences in the target volume
$V(L,R)$ becomes small (of $O(1)$).
Indeed, for the situation $R=0$, where there is only one sequence 
in the target volume to be covered (namely the target sequence $T$),
the estimate (\ref{eq:Nc}) yields $N_c=0$.  For this case, however, 
we can independently estimate the number of randomly situated 
crosstalking sequences required to insure that the target sequence $T$
is covered with probability $1/2$:
\begin{equation}
1-(1-q)^{N_c^{R=0}} = 1/2
\end{equation}
implies
\begin{eqnarray}
N_c^{R=0} &=& \log(1/2) / \log(1-q) \\
&=& \log(1/2) / \log(1-1/V_0)  \label{eq:NcR0}
\end{eqnarray}

The result (\ref{eq:Nc}) assumes an alphabet size $q=2$ (i.e., binary 
sequences).  We can generalize the satisfiability bound in a straightforward
manner, if we assume that binding of two sequences continues to be 
dictated by a maximal Hamming distance, i.e., two sequences $s_1$ and 
$s_2$ will bind if $H(s_1, s_2) \leq R$.  In this case, the form 
of the bound (\ref{eq:Nc})
remains unchanged, and we need simply redefine the relevant sequence volumes 
corresponding to an alphabet of size $q$:
\begin{eqnarray}
V(L,R) = V(L,R,q) &=& \sum_{n=0}^R {L\choose{n}}(q-1)^n  \label{eq:VLRq}\\
V_0(L) = V_0(L,q) &=& q^L   \label{eq:V0Lq}
\end{eqnarray}

In the case of reverse complement symmetric (RCS) sequences (e.g., for
binding of protein to DNA in the regulation of gene transcription), 
the bound is 
reduced because each sequence in the target volume can be covered either by 
a ball centered 
within Hamming distance $R$ of the sequence, or by a ball centered within
distance $R$ of the reverse complement of that sequence.  This has the
effect of doubling the coverage ratio $q$: $q \equiv 2V/V_0$.  As a result,
\begin{equation}
N^{RCS}_c = \frac{{\log(1/V)}}{{\log(1-2V/V_0)}} \label{eq:NcRC}
\end{equation}
which is valid for $R < L/2$.  For $R \geq L/2$, $N^{RCS}_c = 1$.

The main text alludes to a symmetric generalization of the SNQ
that asks whether every protein in a collection is
distinguishable, that is, whether there is a separate sequence niche
for each of $N$ proteins.  While we do not have a general estimate for the 
critical number of proteins $N_c$ for this problem, 
we can produce such an estimate for the special case of $R=0$, where 
crosstalk occurs only if two sequences are exactly the same (no mismatches).
In that limit, the question boils down to this: For binary sequences of 
length $L$, how many randomly chosen sequences must be chosen for there 
to be a probability of at least $1/2$ that two sequences are identical?
This is just the classic ``birthday problem'' of probability theory, 
for a system where a ``year'' contains $V_0 = 2^L$ possible days
(see,
e.g., \verb#http://en.wikipedia.org/wiki/Birthday_problem#).
The probability $p(n)$ that two sequences out of $n$ will match is:
\begin{equation}
p(n) = 1 - \frac{V_0!}{(V_0-n)!\ {V_0}^n} \label{eq:GNcR0}
\end{equation}
so, for a given sequence length $L$,
we can find the number $N_c$ for which this probability exceeds 
$1/2$ to arrive at an estimate for the $R=0$ bound of the generalized SNQ.

In this light, the $R=0$ case for the original SNQ (eq. (\ref{eq:NcR0})) can
be seen as a variant of the ``my birthday problem'', which asks for 
the probability that someone in a group of $N$ people will share \emph{my}
birthday.  The probability of any crosstalk sequence matching the 
target sequence (in the original SNQ) is of course smaller than the 
probability that any two crosstalk sequences will match each other 
(in the generalized SNQ).  For $R>0$, estimating the bound would seem 
to be a variant of the near-match birthday problem \cite{Abramson1970},
but in higher dimensions.

\begin{figure*}[!t]
\centering
\includegraphics[width=120mm]{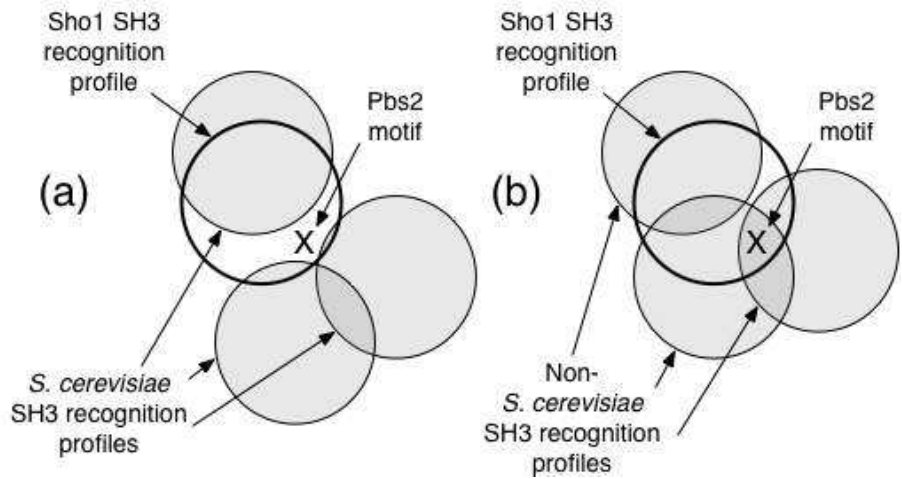}
\caption{\label{Fig_Zarrinpar}
The interpretation offered by 
Zarrinpar, Park and Lim to describe (a) the lack of crosstalk 
among \emph{S. cerevisiae} SH3 domains and (b) the presence of crosstalk among 
non-\emph{S.cerevisiae} SH3 domains. 
[Adapted from \cite{Zarrinpar2003}.]  (a) In 
\emph{S. cerevisiae}, evolutionary selection against crosstalk has driven the 
proline-rich Pbs2 motif to a niche where it is recognized only by the Sho1 SH3
domain.  (b) There is no such selection pressure in other organisms, so 
domains introduced from elsewhere can bind Pbs2.
}
\end{figure*}

\subsection{Size of the largest solution cluster}

Fig.~4(d) of the main text demonstrates that the size
$S_0$ of the largest cluster (solid line) decreases roughly exponentially 
with crosstalk number $N$.  From the geometric argument illustrated in
Fig.~3(a) in the main text, we might expect 
\begin{equation}
S_0 \sim (1-q)^N \approx \exp(-q N) \rm{\ \ \ for\ small\ } q  \label{eq:S0_N}
\end{equation}
where $q \equiv V(L,R)/V_0(L)$ as in eq. (\ref{eq:def_q}).
For $L=16, R=6$, $q \approx 0.23$, and a fit to the
cluster size data in Fig.~4(d) reveals $S_0 \sim \exp(-0.29 N)$.  
The exponential approximation to the power law in eq. (\ref{eq:S0_N}) 
would be more accurate 
for smaller $q$, but part of the discrepancy between the predicted 
and measured decay rate is due to the fact 
that the geometric argument only describes the elimination of viable 
sequences by crosstalk proteins, and not the fragmentation of clusters.
Some of the decrease in $S_0$ is due to the latter effect.

\subsection{Review of results from Zarrinpar, Park and Lim}

We describe here in slightly more detail the experimental results 
of ref. \cite{Zarrinpar2003}.
Zarrinpar \emph{et al.} investigated SH3-mediated signaling in
yeast (\emph{Saccharomyces cerevisiae}), 
probing in particular the signaling pathway
involved in a high-osmolarity response, predicated on the interaction
of the Sho1 protein (containing an SH3 domain) and the Pbs2 protein
(with an exposed proline-rich, PXXP, peptide sequence).
Experimentally, they created chimeric versions of the Sho1 protein, 
replacing the native SH3 domain with each of the other 26 SH3 domains 
found in yeast. (Three of the Sho1 chimeras were insoluble,
however, so they could not be assayed \emph{in vivo}.)
They then sought to determine whether any of those domains could 
reconstitute the function
of the high-osmolarity pathway, and found that none of the other yeast
domains could so function.  \emph{In vitro} peptide binding assays
also carried out revealed a similar lack of interaction from any but
the Sho1-Pbs2 pair.  When SH3
domains from 12 metazoan proteins were tested (both \emph{in vivo}
and \emph{in vitro}), however, it was
discovered that 6 of those were able to reconstitute the function of the
high-osmolarity pathway.  Their interpretation was that there has been
an evolutionary selection against crosstalk in yeast, whereby domains
and peptides have evolved such that the Pbs2 PXXP motif lies
in a niche in sequence space where it is recognized by only the Sho1
SH3 domain, as is illustrated schematically in
Fig.~\ref{Fig_Zarrinpar}(a).  Since there has been no such selection
pressure in other organisms, it was perhaps not surprising that the
Pbs2 motif overlaps with the recognition volumes of many of non-yeast
SH3 proteins, as is illustrated in Fig.~\ref{Fig_Zarrinpar}(b).

Zarrinpar \emph{et al.} also sought to characterize the nature of
protein-protein interactions in the sequence space surrounding the
wild-type Pbs2 motif, which they did by assaying a library of
19 single-base-pair
missense mutations to the native yeast Pbs2 motif (leaving the core 
prolines of the PXXP motif unchanged).  While some mutations resulted 
in increase affinity for Sho1, and some resulted in decreased affinity,
all mutations resulted in an
increased cross-reactivity with other yeast SH3 domains.  This suggests 
that the wild-type Pbs2 is optimized not for affinity, but for discrimination
among different SH3 domains.

\subsection{Methods}

To ascertain whether a given instance of the SNQ was satisfiable or
not, I implemented the algorithm by Gramm \emph{et al.} \cite{Gramm2003}
(``Algorithm D'' in \cite{Gramm2003}, 
modified as described to treat the Distinguishing String
Selection Problem).  This is a recursive, backtracking algorithm in
the style of Davis-Putnam(DP)-type methods used in the study of
other NP-complete problems (e.g., $k-$SAT \cite{Davis1960}). 
Algorithm D in \cite{Gramm2003} implements heuristics to prune 
the search tree, tailored to the 
Distinguishing String Selection Problem (DSSP).
DP-type algorithms are known to be significantly
slower in practice for $k-$SAT than other algorithms (e.g.,
WalkSAT \cite{Selman1993} or survey propagation \cite{Mezard2002}), but
have the advantage of being \emph{complete}, i.e., able to determine
whether any instance is satisfiable or not, given sufficient computer
time.  (Incomplete algorithms can typically find a solution if there
is one, but are not guaranteed to stop if there is no solution.)  For
forays into a newly-identified NP-complete problem such as this,
complete algorithms are a useful first step.  For each SNQ instance,
it was determined whether the instance was satisfiable, and how long 
it took to decide that question.  Since DP-type methods are recursive,
it is conventional to measure algorithm run times in units of number of 
calls to the recursive core, which is what we have done here. 

The SNQ, as stated, applies to any set of sequences $T$ and $\{C\}$.  
This paper has focused on random instances of the SNQ, where the relevant
sequences are sampled uniformly at random from the set of all binary
sequences of length $L$, with equal probabilities of $0$ and $1$ in 
the sequences $T$ and $\{C\}$.  
Simulations of random instances of the SNQ
were carried out, for various values of the relevant control
parameters: the string length $L$, the Hamming radius $R$, and the
number of crosstalk proteins $N$.  Average satisfiability and algorithmic 
run time were computed from 100 random SNQ instances for each set of 
$L$, $R$, and $N$.

To explore the full solution space of SNQ instances, exhaustive
examination was carried out.  For each of the possible $2^L$
sequences, it was determined whether that sequence satisfied the given
SNQ.  The set of valid solutions was assembled to form an undirected
graph, whose nodes were SNQ solutions and whose edges joined nodes
with sequences that differed by Hamming distance of 1, i.e., by 1 bit
flip.  The network analysis package NetworkX [networkx.lanl.gov] was
used to compute connected components of the resulting graphs, and to
generate layouts for visual display.  This work motivated a
contribution on my part to the NetworkX source code repository
[networkx.lanl.gov/changeset/223], using tuples of index coordinates
to label grid graphs, such as would be used to represent an
$L$-dimensional hypercube.  This representation is natural for graphs
connecting nodes in sequence space.  A spring force layout algorithm
was used to generate the images in Figs. 4(a)-(c) in the main text,
whereby connected nodes are attracted to each other to produce compact
representations of connected components.  As noted, however, the
positions of the graph nodes in Figs. 4(a)-(c) have no intrinsic
meaning, as all nodes are vertices on the $L$-dimensional hypercube.
The problem of usefully visualizing complex network structures in
high-dimensional sequence spaces is an ongoing challenge in
computational biology.

\bibliography{myers_arxiv_seqniche}

\end{document}